\documentclass[a4paper]{spie}  %>>> use this instead for A4 paper
\usepackage[]{graphicx}
\usepackage{amssymb,amsmath,psfrag,url}

\newcommand{\Field}[1]{{\boldsymbol{#1}}}

\newcommand{\Kvec}[1]{{\vec{#1}}}

\title{Benchmark of FEM, Waveguide and FDTD Algorithms\\ for Rigorous Mask Simulation} 

\author{
Sven Burger\supit{\,ab}, 
Roderick K\"ohle\supit{\,c}, 
Lin Zschiedrich\supit{\,ab},
Weimin Gao\supit{\,d},\\
Frank Schmidt\supit{\,ab}, 
Reinhard M\"arz\supit{\,be}, 
Christoph N\"olscher\supit{\,f}
\skiplinehalf
\supit{a}
Zuse Institute Berlin,
Takustra{\ss}e 7,
D\,--\,14\,195 Berlin,
Germany\\
DFG Forschungszentrum {\sc Matheon},
Stra{\ss}e des 17.\,Juni 136, 
D\,--\,10\,623 Berlin,
Germany
\smallskip\\
\supit{b}
JCMwave GmbH,
Haarer Stra{\ss}e 14a,
D\,--\,85\,640 Putzbrunn, 
Germany
\smallskip\\
\supit{c}
Infineon Technologies AG,
MP PD CS ATS,\\
Balanstra{\ss}e 73,
D\,--\,81\,541 M\"unchen,
Germany
\smallskip\\
\supit{d}
Sigma-C Software AG,
Thomas-Dehler-Stra{\ss}e 9,
D\,--\,81\,737 M\"unchen,
Germany
\smallskip\\
\supit{e}
Infineon Technologies AG,
COM CAL D TD RETM PI,\\
Balanstra{\ss}e 73,
D\,--\,81\,541 M\"unchen,
Germany
\smallskip\\
\supit{f}
Infineon Technologies SC300 GmbH\&Co OHG,
IFD P300LM,\\
K\"onigsbr\"ucker Stra{\ss}e 180,
D\,--\,01\,099 Dresden,
Germany
}

\authorinfo{
Corresponding author: S. Burger\\
URL: http://www.zib.de/nano-optics/\\
Email: burger@zib.de
}

\begin{document} 
  \maketitle 

%% SPIE Copyright form 
\noindent
Copyright 2005  Society of Photo-Optical Instrumentation Engineers.\\
This paper has been published in Proc.~SPIE {\bf 5992}, pages 378-389
(2005),  
({\it 25th Annual BACUS Symposium on Photomask Technology, 
J.~T.~Weed, P.~M.~Martin, Eds.})
and is made available 
as an electronic reprint with permission of SPIE. 
One print or electronic copy may be made for personal use only. 
Systematic or multiple reproduction, distribution to multiple 
locations via electronic or other means, duplication of any 
material in this paper for a fee or for commercial purposes, 
or modification of the content of the paper are prohibited.
%%%%%%%%%%%%%%%%%%%%%%%%%%%%%%%%%%%%%%%%%%%%%%%%%%%%%%%%%%%%% 
\begin{abstract}
An extremely fast time-harmonic finite element solver developed for the transmission analysis
of photonic crystals was applied to mask simulation problems. The applicability was
proven by examining a set of typical problems and by a benchmarking against two
established methods (FDTD and a differential method) and an analytical example.
The new finite element approach was up to 100$\times$ faster than the 
competing approaches for moderate target accuracies,
and it was the only method which allowed to reach high target 
accuracies.
\end{abstract}


%%%%%%%%%%%%%%%%%%%%%%%%%%%%%%%%%%%%%%%%%%%%%%%%%%%%%%%%%%%%%
\section{Introduction}
The complexity of modern photolithography makes 
extensive simulations indispensable~\cite{Erdmann2004a}. 
Modern lithography simulators include modules describing
illumination, transfer of the optical field through the mask and 
aberrating optical system of the lithographic equipment, the propagation inside
the photoresist, the processes leading to the resist image
and -- in advanced systems -- the etching processes 
leading to the etched image. 
After nearly two decades
of lithography simulation, most of the modules along the simulation 
chain have attained a high degree of maturity.
However, the simulation of light propagation through phase shift masks, 
also applied for the stand-alone analysis of masks and mask tolerances, 
is still challenging in terms of both computational time
and accuracy of the results.

The computation of the print image of a whole chip remains extremely demanding
although approximations, multi-threading and even hardware accelerators 
are applied to reduce the runtime of simulations. 
Rigorous simulations are restricted today to small areas
and even those simulations suffer from the high computational effort. 
At the same time, the progress on the semiconductor roadmap forces the need of 
rigorous 3D simulations, in particular also for alternating and attenuated phase masks.
Experimental investigations of the polarization effects
in Hyper NA immersion lithography~\cite{Teuber2005a} support this assertion. 
Further, the demand to assess the process stability by exploring several dose/defocus conditions
in the process window sharpens the shortage of computational resources.

Keeping this background in mind, we evaluated a frequency-domain 
finite-element method (FEM) solver for Maxwell's equations 
which has been successfully applied to a wide range of 
electromagnetic field computations including
optical and microwave waveguide structures, 
surface plasmons, and nano-structured materials.
In addition, the activity was motivated by a preceding,
successful benchmarking~\cite{Maerz2004a} against MPB,
a widely used and highly sophisticated plane-wave solver
used for the Bloch mode analysis of photonic crystals~\cite{Johnson2001a}. 

In this contribution, the new FEM solver is benchmarked against 
an analytical result which is fairly realistic
for lithography, and against two competing algorithms commonly applied
for the simulation of ''thick'' phase masks.
For the simulation of periodic mask patterns in lithography the most prominent rigorous simulation 
methods include the finite-difference time domain algorithm (FDTD) \cite{Taflove2000a,Erdmann2000a} 
and the modal methods such as the 
differential method~\cite{Petit1980a,Kirchauer1998a} 
or the closely related rigorous coupled wave analysis (RCWA)~\cite{Moharam1988a}. 
The methods differ 
in the way Maxwell's equations are numerically solved and how the boundary conditions of the 
interfaces to the unbound regions above and below the mask are established. 
We will give a brief 
description of the algorithms to give a rough idea about the relevant parameters influencing simulation 
speed and accuracy in the next paragraph. A brief description of FEM will be given in 
Section~\ref{FEMchapter}.
 
The FDTD approach discretizes Maxwell's equations in both time and space and solves the 
scattering problem by simulating the field evolution 
through time until the time-harmonic steady-state solution is reached. 
The interfaces to the 
unbound regions are formed by perfectly matched layers (PML). 
Space and time discretization are interdependent (``magic steps'').
Speed and accuracy of the simulation 
depend on the space and time discretization, the total time period and on 
the PML parameters. 
The differential method describes the propagating fields inside the mask by a plane wave 
expansion. Maxwell's equations are thus converted into a system of linear ordinary differential 
equations (ODEs) relating the scattered waves at the upper and lower mask interfaces. Speed and 
accuracy depend on the number of plane waves and on the resolution used for ODE integration.
 
This paper is organized as follows:
Section~\ref{FEMchapter} introduces the concept of the FEM solver.
Section~\ref{3dchapter} presents several proofs of applicability including
3D simulations, 2D simulations for light scattering off line masks under conical 
incidence, and adaptive refinement of FEM meshes. 
Section~\ref{benchmarkchapter} shows benchmark results of the three different 
solvers for a mask with dense  lines and spaces illuminated at normal incidence. 
Section~\ref{analyticalchapter} verifies the accuracy of the FEM solver by examining a 
closely related toy example offering an anlytical solution. 

\section{Frequency-Domain FEM Analysis for Photomask simulations}
\label{FEMchapter}
This paper considers 
light scattering  off a system which is 
periodic in the $x-$ and $y-$directions and is enclosed by  homogeneous 
substrate (at $z_{sub}$) and superstrate (at $z_{sup}$) 
which are infinite in the $-$, resp.~$+z-$direction. 
Light propagation in the investigated system is governed by Maxwell's equations
where  vanishing densities of free charges and currents are assumed. 
The dielectric coefficient $\varepsilon(\vec{x})$ and the permeability 
$\mu(\vec{x})$ of the considered photomasks are periodic and complex, 
$\varepsilon \left(\vec{x}\right)  =  \varepsilon \left(\vec{x}+\vec{a} \right)$, 
$\mu \left(\vec{x} \right)  =  \mu \left(\vec{x}+\vec{a} \right)$.
Here $\vec{a}$ is any elementary vector of the periodic lattice~\cite{Sakoda2001a}.  
For given primitive lattice vectors 
$\vec{a}_{1}$ and $\vec{a}_{2}$ an elementary cell 
$\Omega\subset\mathbb R^{3}$ is defined as
$\Omega = \left\{\vec{x} \in \mathbb R^{2}\,|\,
x=\alpha_{1}\vec{a}_1+\alpha_{2}\vec{a}_2;
0\leq\alpha_{1},\alpha_{2}<1
\right\}
\times [z_{sub},z_{sup}].$
A time-harmonic ansatz with frequency $\omega$ and magnetic field 
$\Field{H}(\vec{x},t)=e^{-i\omega t}\Field{H}(\vec{x})$ leads to
the following equations for $\Field{H}(\vec{x})$:
\begin{itemize}
\item
The wave equation and the divergence condition for the magnetic field:
\begin{eqnarray}
\label{waveequationH}
\nabla\times\frac{1}{\varepsilon(\vec{x})}\,\nabla\times\Field{H}(\vec{x})
- \omega^2 \mu(\vec{x})\Field{H}(\vec{x}) &=& 0,
\qquad\vec{x}\in\Omega,\\
\label{divconditionH}
\nabla\cdot\mu(\vec{x})\Field{H}(\vec{x}) &=& 0,
\qquad\vec{x}\in\Omega,
\end{eqnarray}
\item
Transparent boundary conditions at the boundaries to the 
substrate (at $z_{sub}$) and superstrate (at $z_{sup}$), $\partial\Omega$,
where $\Field{H}^{in}$ is the incident magnetic field (plane wave 
in this case), and $\vec{n}$ is the normal vector on $\partial\Omega$:
\begin{equation}
\label{tbcH}
	\left(
        \frac{1}{\varepsilon(\vec{x})}\nabla \times (\Field{H} - 
        \Field{H}^{in})
	\right)
	\times \vec{n} = DtN(\Field{H} - 
        \Field{H}^{in}), \qquad \vec{x}\in \partial\Omega.
\end{equation}
The $DtN$ operator (Dirichlet-to-Neumann) is  realized with 
the PML method~\cite{Zschiedrich2005b}. 
This is a generalized formulation of Sommerfeld's radiation condition; it
can be realized alternatively by the Pole condition method~\cite{Hohage03a}.
\item
Periodic boundary conditions for the transverse boundaries, $\partial\Omega$,
governed by Bloch's theorem~\cite{Sakoda2001a}:
\begin{equation}
\label{bloch}
\Field{H}(\vec{x}) = e^{i \Kvec{k}\cdot\vec{x}} \Field{u}(\vec{x}), \qquad
\Field{u}(\vec{x})=\Field{u}(\vec{x}+\vec{a}),
\end{equation}
where the Bloch wavevector $\Kvec{k}\in\mathbb{R}^3$ is defined by the
incoming plane wave $\Field{H}^{in}$.

\end{itemize}

Similar equations are found for the electric field 
$\Field{E}(\vec{x},t)=e^{-i\omega t}\Field{E}(\vec{x})$;
these are treated accordingly.

The finite-element method solves Eqs.~(\ref{waveequationH}) -- (\ref{bloch})
in their weak form, i.e., in an integral representation. 
The computational domain is discretized with triangular (2D)
or tetrahedral (3D) patches. 
The functional spaces are discretized using Nedelec's edge elements, 
which are vectorial functions of polynomial order (typically second order) defined 
on the triangular or tetrahedral patches~\cite{Monk2003a}. 
In a nutshell, FEM can be explained as expanding the field 
corresponding to the exact solution of Equation~(\ref{waveequationH}) in the 
basis given by these elements.
This leads to  a large sparse matrix equation (algebraic problem).
For details on the weak formulation, 
the choice of Bloch-periodic functional spaces,
the FEM discretization, and our implementation of the PML
method we refer to previous works~\cite{Zschiedrich2005b,Burger2005a,Zschiedrich2005a}.
To solve the algebraic problem on a personal computer 
either standard linear algebra decomposition techniques (LU-factorization, e.g.,
package PARDISO~\cite{PARDISO})
or iterative methods~\cite{Deuflhard2003a} are used, depending on problem size.
Due to the use of multi-grid algorithms, the computational time and the memory requirements
grow linearly with the number of unknowns. 

From the users's point of view, the FEM approach presented 
here offers the following advantages:
\begin{itemize}
\item 
The expansion into localized functions 
(shared with any FEM and FD approach) 
is adequate for step index profiles occuring in masks.
\item 
The flexibiltiy of triangulations (shared with any FEM approach) 
allows for the simulation of mask imperfections such as sloped etch profiles
and for adaptive mesh-refinement strategies leading to faster convergence. 
\item 
The frequency domain approach (shared with any PW method) 
is adequate for monochromatic or nearly monochromatic illumination.
\item
Edge elements provide ''built-in'' dielectric boundary conditions
crucial for a high precision simulation of
step index profiles.
\item
The mathematical structure of the algebraic problem allows for the use of very 
efficient numerical solvers, i.e., numerical methods where the computational effort 
grows linearly with the number of unknowns only.
\item
The FEM discretization is characterized by two parameters, the mesh width $h$ 
and the thickness of the PML layer $\rho$.
It is mathematically proven that the FEM approach converges with a fixed convergence 
rate towards the exact solution of Maxwell-type problems for mesh width $h \rightarrow 0,$
and $\rho \rightarrow \infty$.~\cite{Monk2003a,Lassas:98a} 
This allows to easily check whether the attained results can be trusted.
\end{itemize}
These advantages result not only in an increased attainable accuracy, 
but -- via the reduced number of unknowns -- also in a significantly reduced
computational effort at moderate target accuracies  required for
lithography simulation. 
The investigated FEM solver JCMharmony includes adaptive grid refinement, 
higher order, 3D Nedelec elements, advanced periodic and transparent boundary conditions 
and flexible interfaces to the drivers and for postprocessing. 
Typical computation times for 3D problems 
are 30\,seconds for problems with $N\approx 30\,000$ unknowns and 5\,minutes for 
problems with $N\approx 150\,000$ unknowns solved on an
actual standard 64\,bit personal computer ({\it AMD Opteron}).
Typical computation times for 2D problems are given in Table~\ref{bm_table_1}.

\section{Features of the FEM solver}
\label{3dchapter}
The range of applications of the FEM approach was examined by means
of several characteric tasks in mask simulation.
\subsection{Conical Incidence}
\label{oblique_incidence}

\begin{figure}[htb]
\centering
\psfrag{beta}{$\beta$}
\psfrag{w}{\sffamily w}
\psfrag{p}{\sffamily $\mbox{p}_x$}
\psfrag{x}{\sffamily x}
\psfrag{z}{\sffamily z}
\psfrag{h}{\sffamily h}
\psfrag{b}{} %{\it \footnotesize boundary of the computational domain}
\psfrag{air}{\sffamily  air}
\psfrag{substrate}{\sffamily  substrate}
\psfrag{line}{\sffamily  line}
\psfrag{pc}{\sffamily pc}
\includegraphics[width=0.5\textwidth]{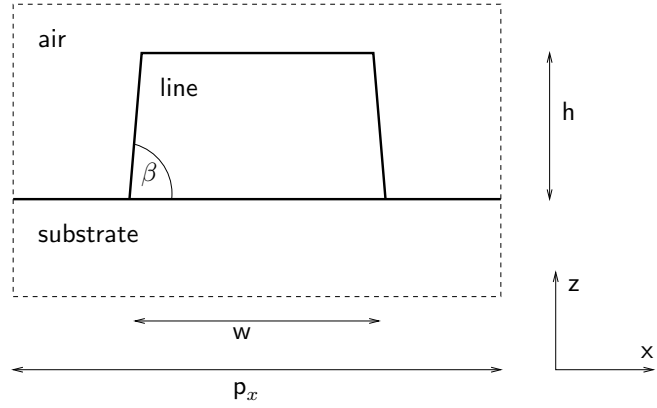}
\caption{
Schematics of the geometry of a periodic linemask:
The computational domain consists of a 
line of width $w$ (at center of the line), height $h$ and  sidewall 
angle $\beta$,  on a  substrate material $SiO_2$, 
surrounded by air.
The geometry is periodic in $x$-direction with a pitch of $p_x$ 
and it is independent on the $y$-coordinate.
The refractive indices of the different present materials are denoted 
by $n_1$ (line), $n_2$ (substrate) and $n_3$ (air), $n_3=1.0$.
}
\label{schema_bm}
\end{figure}

Here we investigate line masks illuminated under conical indicence
(i.e., oblique incidence with respect to both mask plane and grating lines).
This is crucial for the accurate simulation of off-axis source points for dipole, quadrupole or 
annular illumination. 
Figure~\ref{schema_bm} shows the schematics the geometry of
the problem. 
The geometrical, material and source parameters are given in 
Table~\ref{bm_table_2} (data set 1).
The geometry does not depend 
on the $y$-component, therefore Eqn.~(\ref{waveequationH}) reduces 
to a simpler equation where 2D
differential operators act on the 3D electric, resp.~magnetic fields. 
Nevertheless, the problem   
is simulated without any approximations.

\begin{figure}[htb]
\centering
\includegraphics[width=0.5\textwidth]{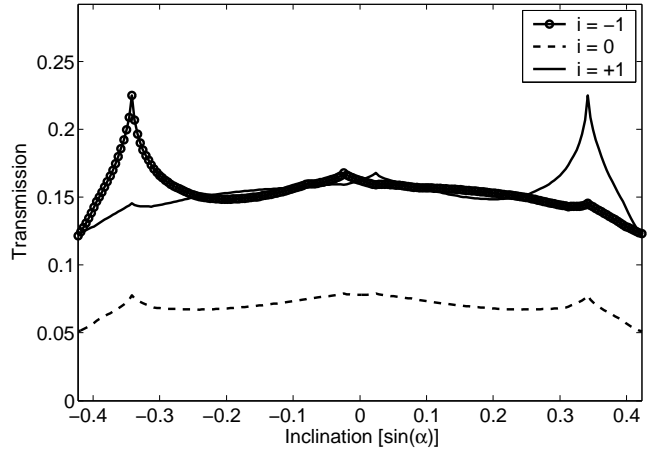}
\caption{Diffraction of a plane wave under conical incidence. 
Intensities of the transmission to the central diffraction orders 
($T=|\vec{A}(\vec{k}_i)|^2, i=-1,0,+1$) in dependence on the 
inclination angle $\alpha$, where $\vec{A}(\vec{k}_i)$ is defined in Eqn.~(\ref{fc_eqn}).
The angle of rotation is constantly $\theta = 20^\circ$.
}
\label{fig_angle_scan}
\end{figure}

\begin{table}[h]
\begin{center}
\begin{tabular}{|l|c|c|c|c|}
\hline
parameter & data set 1 & data set 2 & data set 3 & data set 4 \\ 
\hline 
\hline 
$p_x$ & 400\,nm & 200 -- 800\,nm & 400\,nm & 800\,nm \\ 
$w$   & 200\,nm & $p_x/2$        & 200\,nm & 400\,nm \\ 
\hline 
\hline 
$\beta$  & $86^\circ$                  & \multicolumn{3}{c|}{$90^\circ$} \\ 
$\alpha$ & $-25^\circ$ \dots $+25^\circ$ & \multicolumn{3}{c|}{$0^\circ$} \\ 
$\Theta$ & $20^\circ$                  & \multicolumn{3}{c|}{$0^\circ$} \\ 
\hline 
\hline 
$\lambda_0$ & \multicolumn{4}{c|}{193.0\,nm } \\ 
$h$         & \multicolumn{4}{c|}{65.4\,nm} \\ 
$n_1$       & \multicolumn{4}{c|}{$2.52+0.596i$} \\ 
$n_2$       & \multicolumn{4}{c|}{$1.56306$} \\ 
\hline 
\end{tabular} 

\caption{Parameter settings for the simulations in 
Sections \ref{oblique_incidence} (data set 1), \ref{dop_chapter} (set 2), 
\ref{adaptive_chapter} (set 3)
and \ref{benchmarkchapter} (set 4). 
}
\label{bm_table_2}
\end{center}
\end{table}

The performance of the FEM solver is demonstrated by 
a parameter scan for varied inclination angle $\alpha$ of 
the source ($S$-polarization).
The wavevector $\vec{k}$ of the incident plane wave is attained by a 
rotation of the vector $(0,0,2\pi/\lambda)$ (where $\lambda=\lambda_0/n_2$, 
vacuum wavelength $\lambda_0$, refractive index $n_2$) around the $y$-axis by 
the inclination angle $\alpha$
and a subsequent rotation around the $z$-axis by the rotation angle $\Theta$.
In this scan we fix the rotation, $\Theta=20^{\circ}$ and vary the inclination, 
$\alpha=-25\dots 25$, further 
parameters are given in Table~\ref{bm_table_2}. This 
yields an incident wave vector which is scanned from 
$\vec{k}\approx (-2.021,-0.736,4.612)10^7/$m to $\vec{k}\approx (2.021,0.736,4.612)10^7/$m. %,
Figure~\ref{fig_angle_scan} shows the normalized magnitude of the Fourier coefficients corresponding to the 
zero and first diffraction orders of the scattered light field in dependence on the angle of incidence.
A typical computation time for a single data point in this scan was 15\,sec ($N\approx 3\times10^4$\,unknowns,
adaptive grid refinement, computation on a personal computer/Intel Pentium IV, 2.5\,GHz), resulting in 
a total time of roughly 1\,h for the scan with 200 data points.

\subsection{Degree of Polarization of Light Transmitted through a Line Mask}
\label{dop_chapter}

We have performed a scan over different geometrical parameters by varying the pitch and 
the linewidth of a line-mask. 
Geometrical, material and source parameters are again given in Fig.~\ref{schema_bm}
and in Table~\ref{bm_table_2} (data set 2).
Since the plane of incidence is normal to the grating lines ($k_y = 0$),
TE- and TM-polarization is supported, i.e., the problem becomes scalar.

\begin{figure}[htb]
\centering
{\sffamily a)}\includegraphics[width=0.475\textwidth]{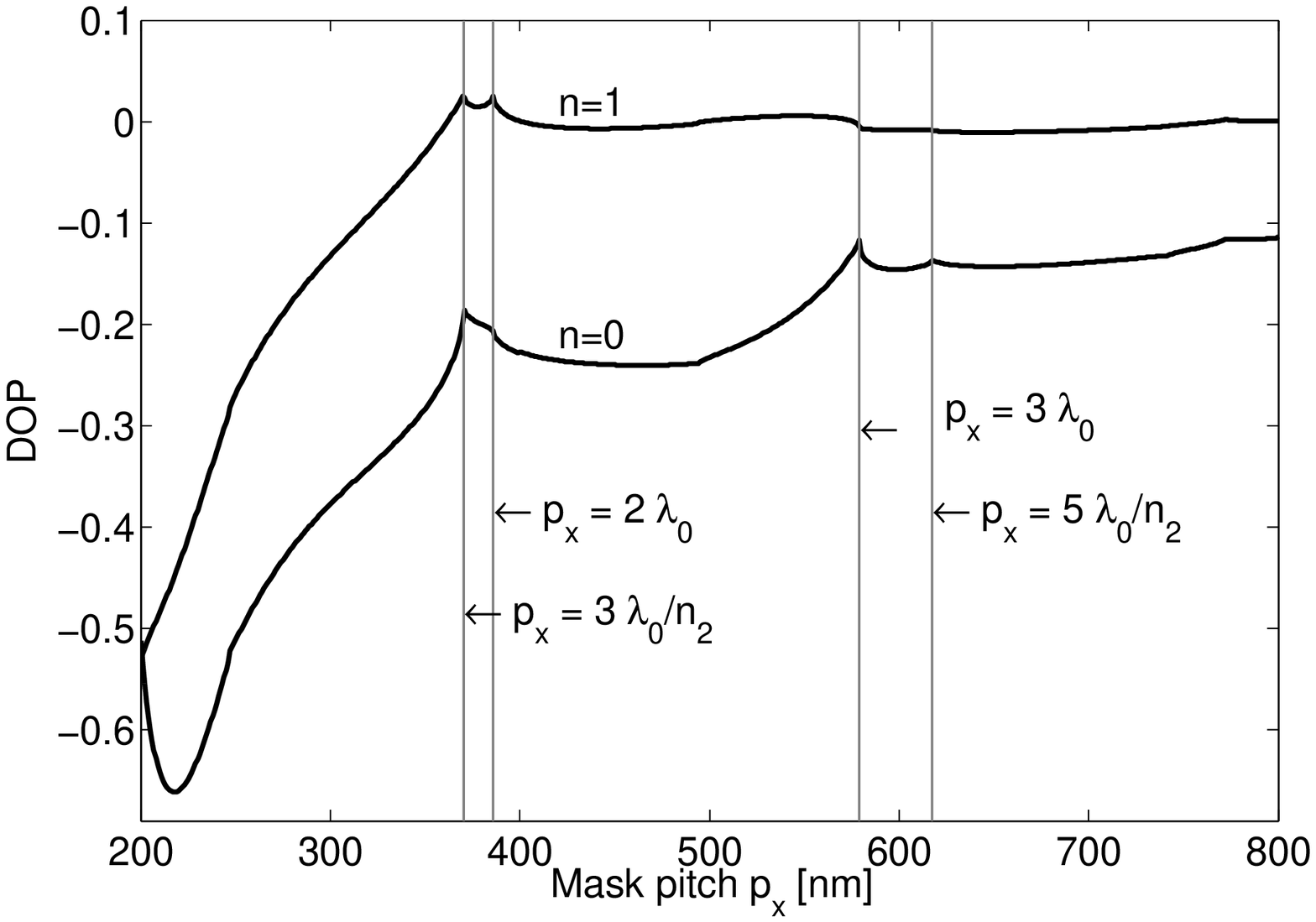}
{\sffamily b)}\includegraphics[width=0.475\textwidth]{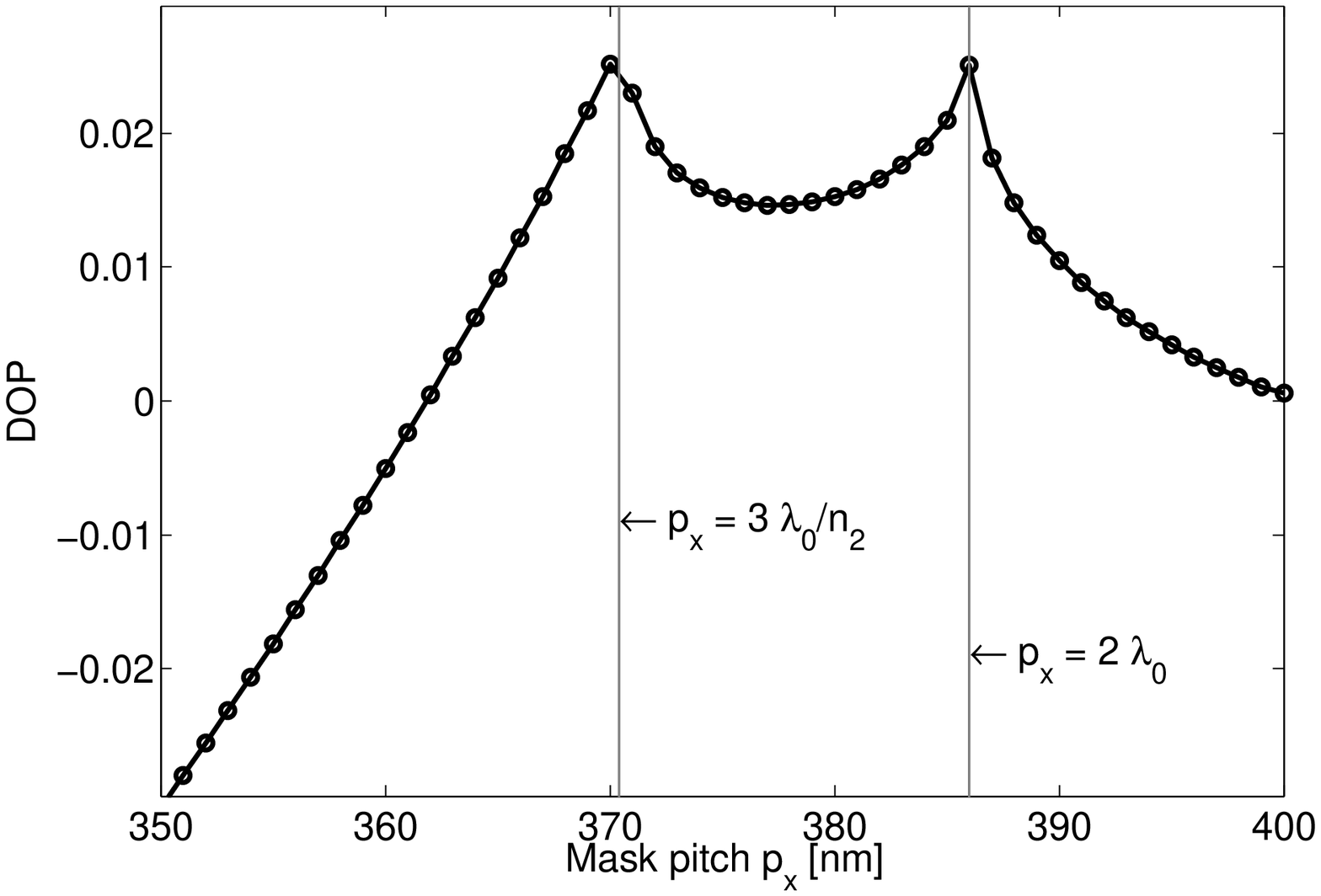}
\caption{(a) Degree of polarization in dependence on mask pitch ($p_x$) for light diffracted 
to the zero ($i=0$) and first ($i=1$) diffraction order.
Several strong Wood's anomalies are indicated at $p_x=N\lambda$.
(b) Enlargement of a detail.}
\label{fig_dop_scan}
\end{figure}
	
Figure~\ref{fig_dop_scan} shows the degree of polarization of the zero and first 
transmitted diffraction orders, defined as 
$DOP=(I_{TE}-I_{TM})/(I_{TE}+I_{TM})$ in dependence on the pitch $p_x$.
The strong Wood anomalies~\cite{Petit1980a}, 
some of which are also verified by Teuber et al.~\cite{Teuber2005a},
are caused by the excitation of waves traveling along the 
mask surface.
We have constructed transparent boundary conditions for 
the whole range of investigated pitches, including 'regular' regions where 
the transmitted diffraction orders correspond to plane waves with a nonzero 
$z$-component of the wavevector and regions close to Wood's anomalies 
where certain diffraction orders cannot propagate as plane waves anymore.
We are currently implementing an adaptive strategy for the PML implementation 
of the transparent boundary conditions in order to automatically account for 
such effects. 

The average computation time for a single data point in this scan was about 3.5\,sec 
($N\approx 1 - 4\times10^4$\,unknowns, depending on geometry size, yielding a 
relative error of the diffraction intensities of less than 1\%;
computation on a personal computer/Intel Pentium IV, 2.5\,GHz), resulting in 
a total time of roughly 75\,min for the scan with 1200 data points.
Similar results are obtained with the solvers SOLID E and Delight 
(see Section~\ref{benchmarkchapter}), 
however, the Wood's anomalies are not or less accurately resolved.

\subsection{Adaptive Grid Refinement}
\label{adaptive_chapter}

By refining the resolution of the geometry-triangulation 
the accuracy of the solution is increased. 
FEM-solvers use as a standard a {\it regular} grid refinement, 
i.e., in 3D {\it each} tetrahedron of the 
discretization is refined to eight smaller tetrahedra,
in 2D {\it each} triangle of the discretization is refined to four smaller triangles.
However, FEM meshes also allow for {\it adaptive} strategies
where only certain elements of the triangulation are refined. 
The investigated solver JCMharmony uses a residuum-based 
error-estimator~\cite{Heuveline2001a} for adaptive grid refinement.
Obviously, adaptive grid refinement is especially useful when the 
sought solutions are geometrically localized, or
when the geometry exhibits sharp features, like discontinuities in the
refractive index distribution.

\begin{figure}[htb]
\centering
\includegraphics[width=0.5\textwidth]{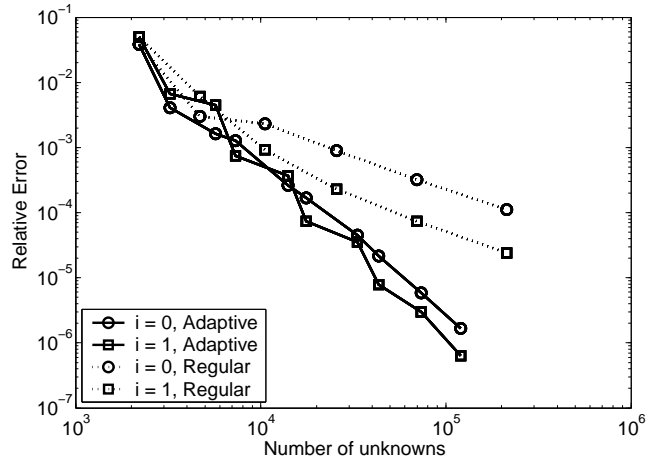}
\caption{Comparison adaptive vs.~regular FEM grid refinement.
Relative error of the transmission intensity in the 
zero (circles) and first (squares) order 
of TM-polarized light incident onto a periodic line mask, calculated with the FEM solver 
JCMharmony.
Solid lines correspond to automatic adaptive refinement, 
dotted lines correspond to regular refinement.}
\label{fig_adaptivity}
\end{figure}

We compare the different refinement strategies by 
observing the convergence of the solutions obtained for increased 
grid resolution (i.e., increased number of unknowns). 
The setting is similar to the one in Section~\ref{dop_chapter}, the parameters are 
given in Table~\ref{bm_table_2} (data set 3), and the incident light is TM-polarized.
Fig.~\ref{fig_adaptivity} shows the convergence of the relative error 
of the light intensity, $\Delta I=|I_{N,i}-I_{\inf,i}|/I_{\inf,i}$, 
in two different diffraction orders for 
adaptive  and for regular grid refinement. 
Here, $I_{N,i}$ denotes the light intensity in the $i^{th}$ diffraction order calculated 
from a solution with $N$ unknowns, $I_{\inf,i}$ denotes the intensity calculated on a very fine 
grid ({\it quasi-exact solution}).
In this example, the use of the error estimator and adaptive refinement yields 
two orders of magnitude in the accuracy of the error for a number of unknowns of 
$N\sim 10^{5}$.

\subsection{Fully 3D Simulations}

\begin{figure}[htb]
\centering
\psfrag{x}{\sffamily x}
\psfrag{y}{\sffamily y}
\psfrag{z}{\sffamily z}
\psfrag{a}{\sffamily a)}
\psfrag{b}{\sffamily b)}
\psfrag{c}{\sffamily c)}
\psfrag{d}{\sffamily d)}
\psfrag{e}{\sffamily e)}
\psfrag{f}{\sffamily f)}
\includegraphics[width=0.95\textwidth]{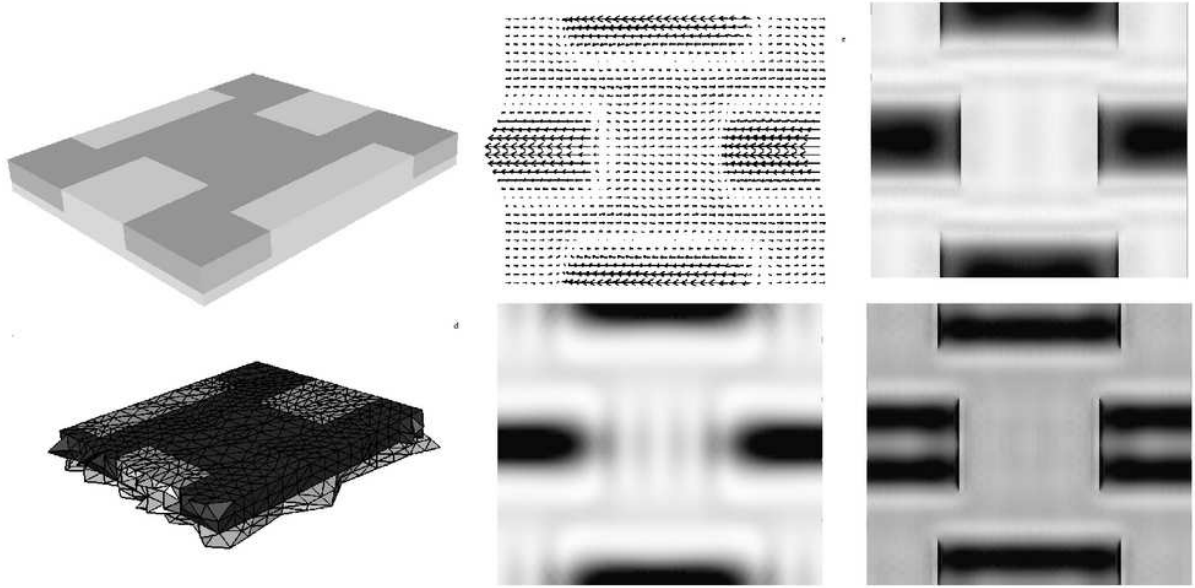}
\caption{Unit cell ($920\,{\rm nm} \times 800\,{\rm nm}  \times 85.4\,{\rm nm}$) 
of a periodic mask pattern 
(a) built up from absorbing material (dark gray, $n_1$) and air (light gray)
and its initial tetrahedral triangulation (b).
The mask is illuminated at normal incidence.   
The graphs (c) and (d) show the (projected) vectorial solution and 
the corresponding intensity gray scale map on a cross section well below the output
side of the mask, (e) and (f) show intensity maps for cross sections 
in the middle and at the 
input side of the mask, respectively (white: low intensity, black: high intensity).
{\footnotesize (See original publication for images with higher resolution.)}
}
\label{fig_3d}
\end{figure}

As a true 3D example, we have examined the transmission through 
a mask with a periodic 3D pattern as shown in Figure 6a (``chequerboard pattern'').
It is illuminated by a plane wave incident from top.
The parameters $h$, $n_1$, $n_2$, $n_3$,
and $\lambda_0$ are the same as in the previous examples.
The mask is discretized by a tetrahedral mesh
supporting second order Nedelec elements~\cite{Monk2003a}. 
The inital triangulation is shown in Figure 6b.
After two refinement steps, the discretization led to a 
linear system with $2.7 \cdot 10^6$ unknows. 
A cross section of the 3D vectorial solution (c) and cross sections through 
the 3D intensity distribution (d-f) are also shown in Fig.~\ref{fig_3d}.
One cleary observes the expected discontinuous behavior of the electric 
field at material interfaces. 
For more details on 3D FEM computations we refer to a previous work~\cite{Burger2005w}.   

\section{Benchmark of different rigorous methods}
\label{benchmarkchapter}
\begin{table}[h]
\begin{center}
\begin{footnotesize}
\begin{tabular}{|rlll|rlll|}
\hline
\multicolumn{4}{|l|}{JCMharmony (regular refinement, TE)} &
\multicolumn{4}{l|}{JCMharmony (regular refinement, TM)} \\
\hline
t[units] & $\Re (FC_0)$ & $\Im (FC_0)$ & $|FC_0|$ &
t[units] & $\Re (FC_0)$ & $\Im (FC_0)$ & $|FC_0|$ \\ 
  5.9  &{\bf -0.1}6969573 &{\bf 0.2}4799014 &{\bf 0.3}0049251 &   5.9  &{\bf -0.2}2884477 &{\bf 0.2}6037940 &{\bf 0.34}665164  \\ 
  8.0  &{\bf -0.157}84440 &{\bf 0.26}773822 &{\bf 0.31}080317 &   8.0  &{\bf -0.2}2041911 &{\bf 0.271}07079 &{\bf 0.349}37652  \\ 
 14.2  &{\bf -0.157}51644 &{\bf 0.269}27004 &{\bf 0.31}195799 &  14.2  &{\bf -0.219}64384 &{\bf 0.2718}9080 &{\bf 0.349}52543  \\ 
 30.7  &{\bf -0.15749}704 &{\bf 0.2693}7915 &{\bf 0.31204}238 &  30.7  &{\bf -0.219}50479 &{\bf 0.2718}4885 &{\bf 0.349}40542  \\ 
 83.9  &{\bf -0.157495}76 &{\bf 0.269386}18 &{\bf 0.31204}780 &  84.5  &{\bf -0.2194}6245 &{\bf 0.2718}1601 &{\bf 0.3493}5327  \\ 
238.4  &{\bf -0.157495}68 &{\bf 0.269386}62 &{\bf 0.31204}814 & 239.2  &{\bf -0.2194}4684 &{\bf 0.2718}0433 &{\bf 0.3493}3438  \\ 
\hline
\hline
\multicolumn{4}{|l|}{JCMharmony (adaptive refinement, TE)} &
\multicolumn{4}{l|}{JCMharmony (adaptive refinement, TM)} \\
\hline
t[units] & $\Re (FC_0)$ & $\Im (FC_0)$ & $|FC_0|$ &
t[units] & $\Re (FC_0)$ & $\Im (FC_0)$ & $|FC_0|$ \\ 
  6.1  &{\bf -0.1}6635461 &{\bf 0.2}5218040 &{\bf 0.3}0210728 &   6.5  &{\bf -0.2}2657994 &{\bf 0.2}6033139 &{\bf 0.34}512447  \\ 
  8.7  &{\bf -0.15}969388 &{\bf 0.26}077231 &{\bf 0.3}0578478 &   9.2  &{\bf -0.2}2010390 &{\bf 0.2}6697463 &{\bf 0.34}600748  \\ 
 13.7  &{\bf -0.157}33135 &{\bf 0.26}825170 &{\bf 0.31}098574 &  14.6  &{\bf -0.2}2002495 &{\bf 0.27}095147 &{\bf 0.34}903536  \\ 
 21.4  &{\bf -0.157}53513 &{\bf 0.269}04719 &{\bf 0.31}177509 &  22.8  &{\bf -0.219}43271 &{\bf 0.271}37278 &{\bf 0.34}898982  \\ 
 39.1  &{\bf -0.1574}8778 &{\bf 0.2693}1712 &{\bf 0.31}198416 &  38.6  &{\bf -0.219}50100 &{\bf 0.2717}5820 &{\bf 0.3493}3252  \\ 
 63.3  &{\bf -0.15749}761 &{\bf 0.2693}7208 &{\bf 0.3120}3656 &  63.4  &{\bf -0.2194}4800 &{\bf 0.2717}7929 &{\bf 0.3493}1563  \\ 
112.7  &{\bf -0.15749}472 &{\bf 0.26938}188 &{\bf 0.31204}356 & 109.5  &{\bf -0.2194}4616 &{\bf 0.27179}933 &{\bf 0.3493}3006  \\ 
185.5  &{\bf -0.157495}85 &{\bf 0.26938}548 &{\bf 0.31204}724 & 181.8  &{\bf -0.21943}960 &{\bf 0.27179}803 &{\bf 0.34932}493  \\ 
323.9  &{\bf -0.1574956}0 &{\bf 0.269386}35 &{\bf 0.31204}786 & 312.7  &{\bf -0.21943}846 &{\bf 0.27179}904 &{\bf 0.34932}501  \\ 
549.7  &{\bf -0.1574956}9 &{\bf 0.269386}57 &{\bf 0.31204}810 & 516.5  &{\bf -0.21943}773 &{\bf 0.27179}869 &{\bf 0.34932}427  \\ 
\hline \hline
\multicolumn{4}{|l|}{SOLID E (TE)} &
\multicolumn{4} {l|}{SOLID E (TM)} \\
\hline
t[units] & $\Re (FC_0)$ & $\Im (FC_0)$ & $|FC_0|$ & 
t[units] & $\Re (FC_0)$ & $\Im (FC_0)$ & $|FC_0|$ \\ 
  7.8  &{\bf -0.2}582035 &{\bf 0.}2073675 &{\bf 0.3}311651 &   9.1  &{\bf -0.2}342025 &{\bf 0.2}523403 &{\bf 0.3}442767  \\ 
 13.9  &{\bf -0.2}528993 &{\bf 0.1}670695 &{\bf 0.30}31011 &  13.5  &{\bf -0.2}512106 &{\bf 0.2}257843 &{\bf 0.33}77652  \\ 
 39.8  &{\bf -0.2}561597 &{\bf 0.1}580582 &{\bf 0.30}09986 &  38.6  &{\bf -0.2}586513 &{\bf 0.2}177905 &{\bf 0.33}81319  \\ 
192.3  &{\bf -0.2}582230 &{\bf 0.1}603459 &{\bf 0.30}39571 & 170.2  &{\bf -0.2}572943 &{\bf 0.2}210333 &{\bf 0.33}91991  \\ 
1178.7 &{\bf -0.26}07888 &{\bf 0.15}37159 &{\bf 0.30}27199 & 1103.6 &{\bf -0.26}22736 &{\bf 0.21}31887 &{\bf 0.33}79894  \\ 
2109.6 &{\bf -0.26}10307 &{\bf 0.15}52241 &{\bf 0.303}6965 & 1938.8 &{\bf -0.26}09029 &{\bf 0.21}43467 &{\bf 0.33}76608  \\ 
12169.9&{\bf -0.26}21858 &{\bf 0.15}37540 &{\bf 0.303}9435 & 11785.0&{\bf -0.26}18512 &{\bf 0.21}07800 &{\bf 0.33}61462  \\ 
\hline \hline
\multicolumn{4}{|l|}{Delight (TE)} &
\multicolumn{4} {l|}{Delight (TM)} \\
\hline
t[units] & $\Re (FC_0)$ & $\Im (FC_0)$ & $|FC_0|$ & 
 & $\Re (FC_0)$ & $\Im (FC_0)$ & $|FC_0|$ \\ 
   2.9  &{\bf -0.1}6846924 &{\bf 0.2}6716937 &{\bf 0.31}585022 &   &{\bf -0.2}1385244 &{\bf 0.2}4047966 &{\bf 0.3}2181258  \\ 
   4.0  &{\bf -0.1}6292394 &{\bf 0.27}139397 &{\bf 0.31}654209 &   &{\bf -0.2}1239923 &{\bf 0.2}4108101 &{\bf 0.3}2129968  \\ 
   5.6  &{\bf -0.15}993653 &{\bf 0.27}081034 &{\bf 0.31}451221 &   &{\bf -0.2}1631258 &{\bf 0.2}4383531 &{\bf 0.3}2595520  \\ 
  12.0  &{\bf -0.15}843702 &{\bf 0.27}004071 &{\bf 0.31}308828 &   &{\bf -0.2}1957244 &{\bf 0.2}5051824 &{\bf 0.3}3312377  \\ 
  23.4  &{\bf -0.15}802117 &{\bf 0.269}80806 &{\bf 0.312}67728 &   &{\bf -0.220}51428 &{\bf 0.2}5495307 &{\bf 0.3}3708696  \\ 
  44.4  &{\bf -0.157}84463 &{\bf 0.269}71301 &{\bf 0.312}50606 &   &{\bf -0.220}81221 &{\bf 0.2}5792098 &{\bf 0.3}3953095  \\ 
 124.3  &{\bf -0.157}70936 &{\bf 0.269}64098 &{\bf 0.312}37558 &   &{\bf -0.220}88603 &{\bf 0.26}158676 &{\bf 0.34}237154  \\ 
 587.0  &{\bf -0.1576}4073 &{\bf 0.269}60489 &{\bf 0.312}30977 &   &{\bf -0.220}67944 &{\bf 0.26}517225 &{\bf 0.34}498657  \\ 
1800.2  &{\bf -0.1576}2403 &{\bf 0.269}59612 &{\bf 0.312}29378 &   &{\bf -0.220}48563 &{\bf 0.26}692949 &{\bf 0.34}621564  \\ 
\hline
\end{tabular}
\caption{Computation times, real and imaginary parts and magnitudes  of the 
$0^{th}$ Fourier coefficients $A_{y}(\vec{k}_{FC}=0)$ for polarizations 
TE and TM and for the three benchmarked methods. Geometrical and material parameters are
denoted in Table~\ref{bm_table_2} (data set 4).
For each method, increased computation time corresponds to a higher spatial resolution.
Please note that with Delight, both, TE and TM modes are computed simultaneously,
JCMharmony was used with regular and with adaptive grid refinement.
Units of the Fourier coefficients are [V/m] (TE), resp. [A/m] (TM).}
\label{bm_table_1}
\end{footnotesize}
\end{center}
\end{table}

We have performed a benchmark of the previously described FEM solver JCMharmony
and two other advanced methods, which are also commercially available:
The Finite-Difference Time-Domain solver SOLID E~\cite{Kalus2001a}, and the 
solver Delight~\cite{Kirchauer1998a}, which relies on a waveguide-method.
In this benchmark we have investigated light propagation through 
a phase mask.
The geometry of the problem is outlined schematically in Figure~\ref{schema_bm}.
A plane electromagnetic wave 
is incident onto the computational 
domain with a wavelength of $\lambda$, a 
wavevector of $\vec{k}=(0,0,+2\pi/\lambda)$ and 
a polarization of $\vec{H}=(0,H_y,0)$, 
with $H_y=n_2^{1/2}\cdot 1\,\mbox{A}/\mbox{m}$  (TM), 
resp.~$\vec{E}=(0,E_y,0)$, $E_y=n_2^{-1/2}\cdot 1\,\mbox{V}/\mbox{m}$ (TE).
The geometrical and material parameters are denoted in 
Table~\ref{bm_table_2} (data set 4). 

For a quantitative assessment of the different simulation methods we 
investigate the internal convergence behavior of the simulation results, i.e.,
the deviation of the simulation results from the highest effort result
obtained with the same method. 
It turnes out that in all cases the accuracy of the field representation 
inside the computational domain dominates the convergence speed. 
For FDTD both space and time resolution are affected since the stability of the 
method requires aligning time resolution with spatial resolution. 
For the differential method, the spatial resolution is determined by the number of 
Fourier coefficients -- 
since the grating considered here exhibits discontinuities in the material distribution, evanescent 
waves play an important role for correctly approximating the fields in the mask. 
For the FEM solver, 
the accuracy of the solution depends on the number of unknowns which is given by the number of 
geometrical patches (triangles in 2D) and 
by the parameters of the polynomial functions defined on the finite elements. 
The focus of the benchmarking is to compare the convergence speed 
rather than absolute runtime, i.e., the accuracy gain obtained by an increase in runtime. 
Therefore, other settings of the simulators were conservatively 
chosen in order to avoid any influence on the accuracy. 
This means, we expect that a more aggressive tuning of these parameters or 
implementations might slightly reduce the absolute runtime for all methods.

As output we monitor the coefficients of the Fourier decomposition of 
the solution at the output boundary of the computational domain. 
The square of the Fourier coefficients is proportional to the 
power of light diffracted into the corresponding diffraction order 
of the periodic mask.

The Fourier coefficient of the $y$-component of the investigated 
field $\vec{f}=\vec{E}\mbox{, resp. }\vec{H}$ is defined as 
\begin{equation}
A_{y}(\vec{k}_{FC})=
\frac{1}{p_x}
\int_{-p_x/2}^{p_x/2}
f_y(x,y,z_{0})
\exp(-i \vec{k}_{FC}\vec{x} )  dx\, ,
\label{fc_eqn}
\end{equation}

where $\vec{k}_{FC}$ is the projection of the wavevector of the investigated diffraction order
onto the  $x-y-$plane ($\vec{k}_{FC}=0$ for zero order and perpendicular incidence). 
Table~\ref{bm_table_1} lists the Fourier coefficients for TE and TM polarization, 
obtained from the solutions of the 
the three benchmarked solvers with different resolutions.
JCMharmony has been used in adaptive and in regular grid refinement mode (see 
Section~\ref{adaptive_chapter}).
To ease the use of Table~\ref{bm_table_1}, we have marked the already converged digits in bold.

\begin{figure}[htb]
\centering
{\sffamily a)}\includegraphics[width=0.475\textwidth]{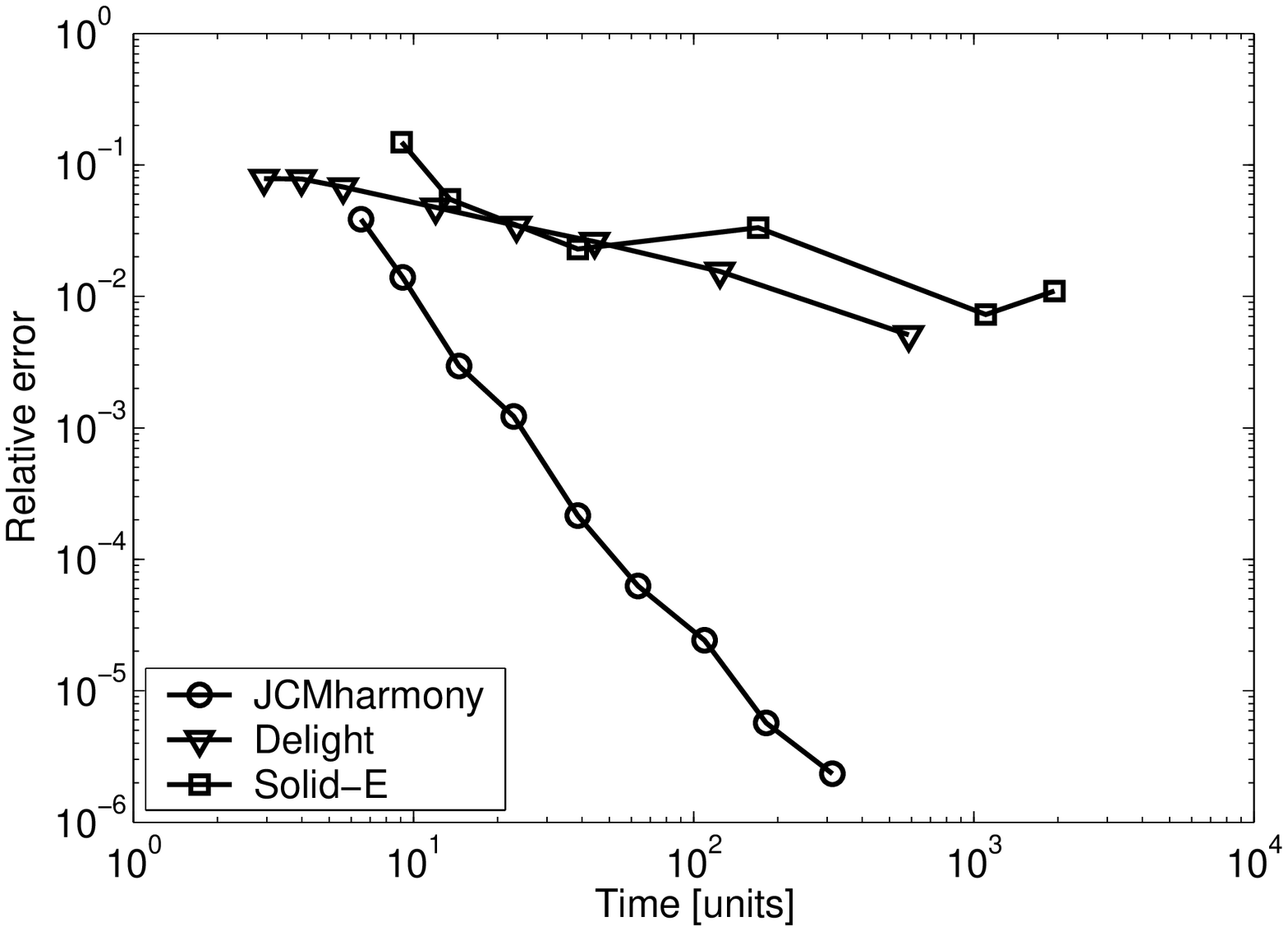}
{\sffamily b)}\includegraphics[width=0.475\textwidth]{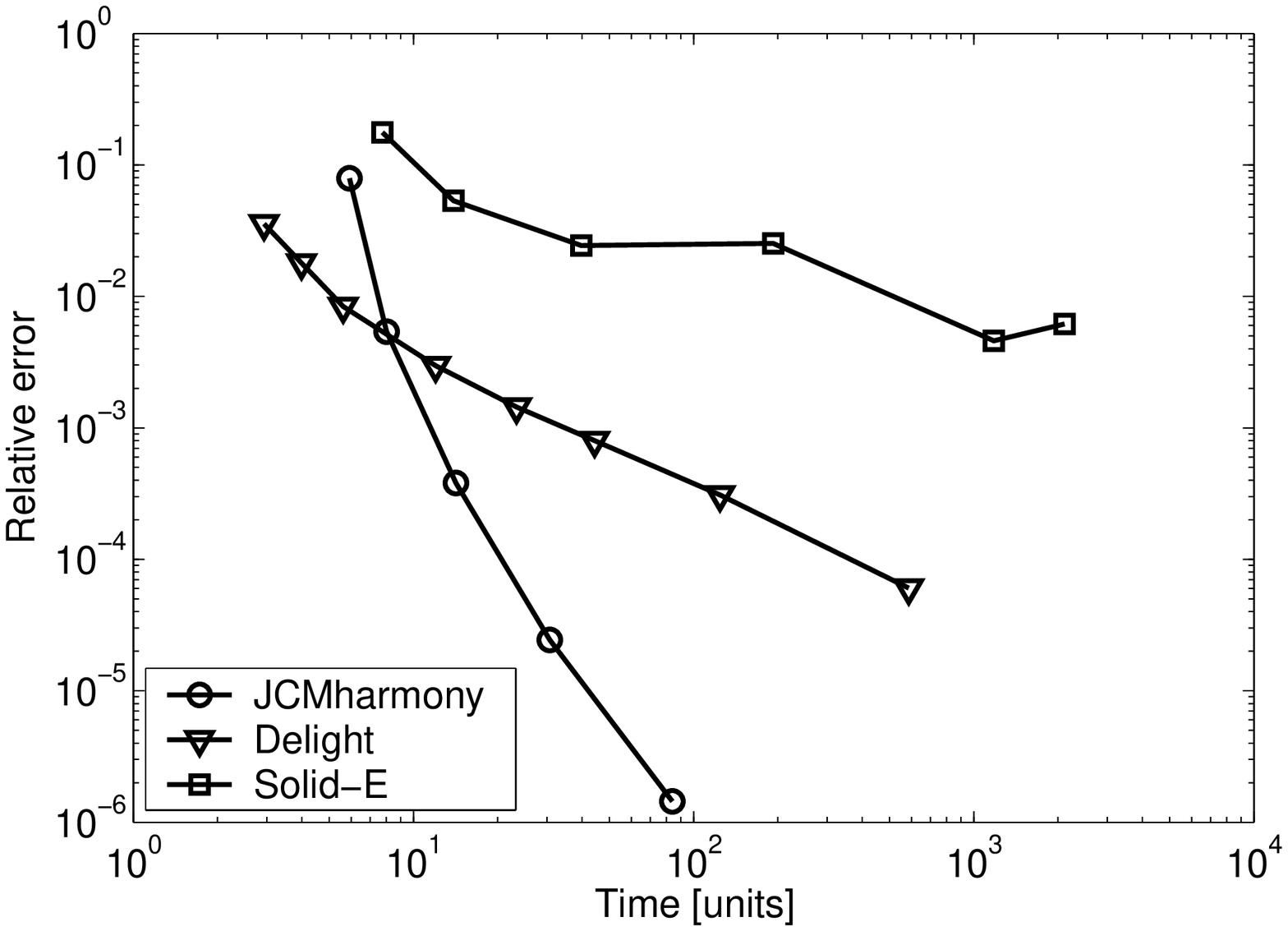}
\caption{
Relative error ($\Delta A_0$) of the $0^{th}$ complex 
Fourier coefficient vs.~normalized computation time for TM-polarization (a) and 
TE-polarization (b).
Circles correspond to results obtained with JCMharmony (with adaptive (a), resp.~regular (b) 
grid refinement mode), triangles to Delight and squares to SOLID E.
The corresponding data is listed in Table~\ref{bm_table_1}.
}
\label{conv_800_tm_te}
\end{figure}

Figure~\ref{conv_800_tm_te} shows the convergence 
of the different methods. 
Plotted is the relative error of the $0^{th}$ order complex Fourier coefficient, $A_y(0)$, of 
the simulated field components, 
\begin{equation}
\Delta A_{0}=\frac{|A_y(0)-\tilde{A}_y(0)|}{|\tilde{A}_y(0)|}\,\,,
\label{EqnDeltaA0}
\end{equation}
where $\tilde A_y(0)$ denotes the complex Fourier coefficient computed at highest spatial 
resolution with the corresponding method, vs.~computation time. 
Since the computations were carried out
on different platforms, computation time is given in units of a Matlab FFT run on the
same platform.
One unit of time corresponds to approx.~0.25\,sec on a personal computer (Intel Pentium IV, 2.5\,GHz). 
In spite of the respective internal convergence, we observe that the intensities of the 
$0^{th}$ order diffraction 
computed with the three methods differ significantly ($>2 \%$), especially for TM polarization.
For that reason, we benchmarked the FEM solver using an example providing an analytical solution
(see Section~\ref{analyticalchapter}).  
The speed of convergence of the three methods differs also significantly.

It can be seen from Fig.~\ref{conv_800_tm_te} (or, alternatively, 
Table~\ref{bm_table_2}) that the convergence behavior of the benchmarked solvers 
differs for TE and TM polarization.
In the case of TM polarization, the FEM solver JCMharmony is the only one 
to reach target accuracies $\Delta A_0<10^{-3}$. 
For a low target accuracy of 
$\Delta A_0 > 10^{-2}$ computation times are comparable for all three solvers, 
for a moderate target accurracy of $\Delta A_0 \approx 2 \cdot 10^{-3}$ JCMharmony 
is about 100$\times$ faster than Delight.
SOLID E reaches an accuracy of about $\Delta A_0 \approx 10^{-2}$.
For low resolutions (computation time $< 10$~units) the loading of 
the code requires a significant amount of the total computation 
time (for SOLID E also the loading of the GUI which takes about 6 
time units).
In the case of TE polarization the convergence behavior of Delight is 
better than in the TM case, being slightly faster than JCMharmony at low target accuracies 
 $\Delta A_0>1\%$, comparable at intermediate target accuracies  $\Delta A_0\approx1\%$,
and about 20$\times$ slower than JCMharmony at high target accuracies of 
$\Delta A_0 \approx  10^{-4}$.
Please note that in the TM case 
adaptive grid refinement yields a gain in convergence for the 
FEM solver JCMharmony, while in the TE case it is not superior 
to regular grid refinement (see also Section~\ref{adaptive_chapter}).

\section{Benchmarking with an Analytical Solution}
\label{analyticalchapter}
\begin{figure}[bhtp]
  \centering \includegraphics[width =
  0.45\textwidth]{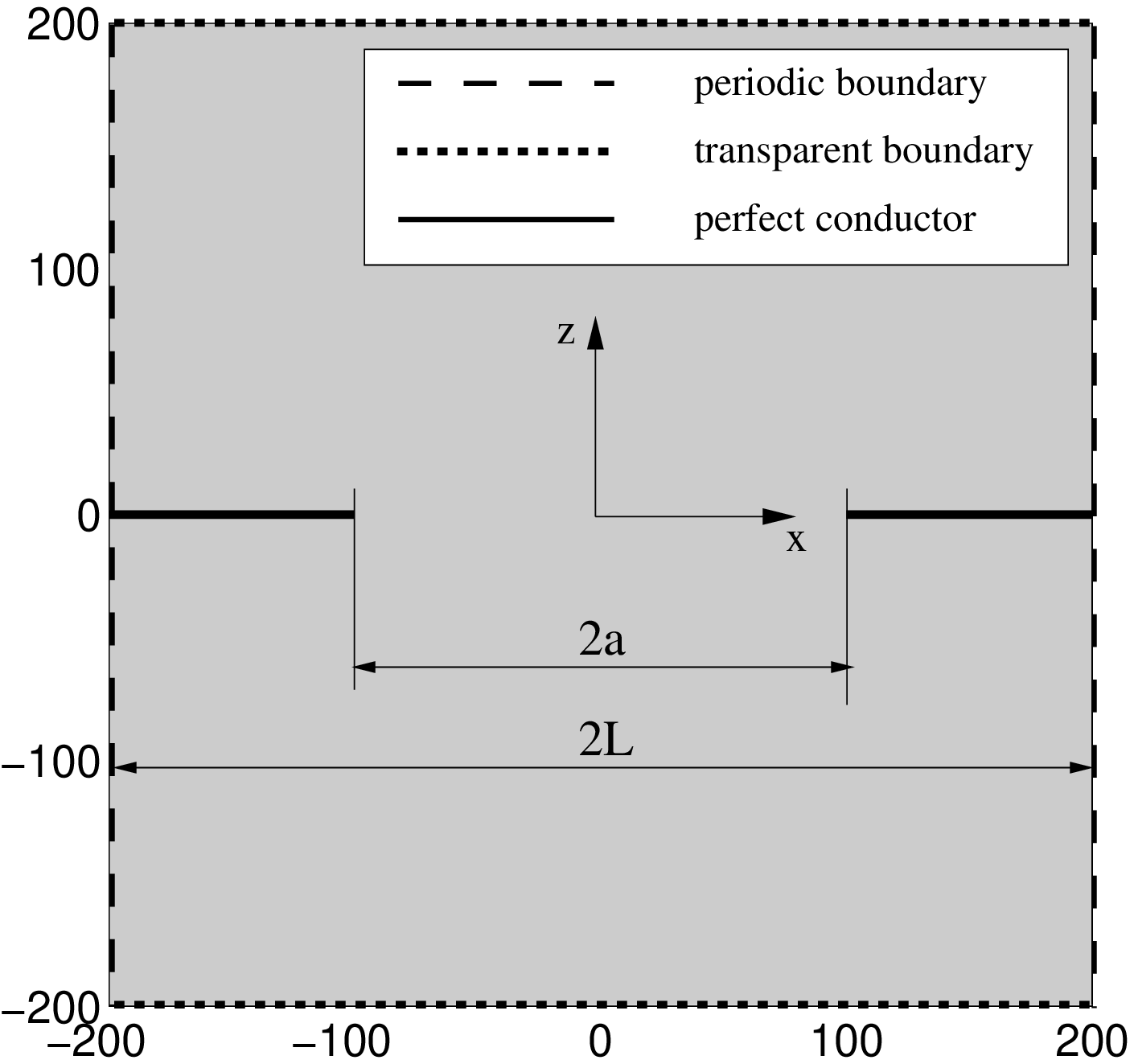}
\includegraphics[width = 0.45\textwidth]{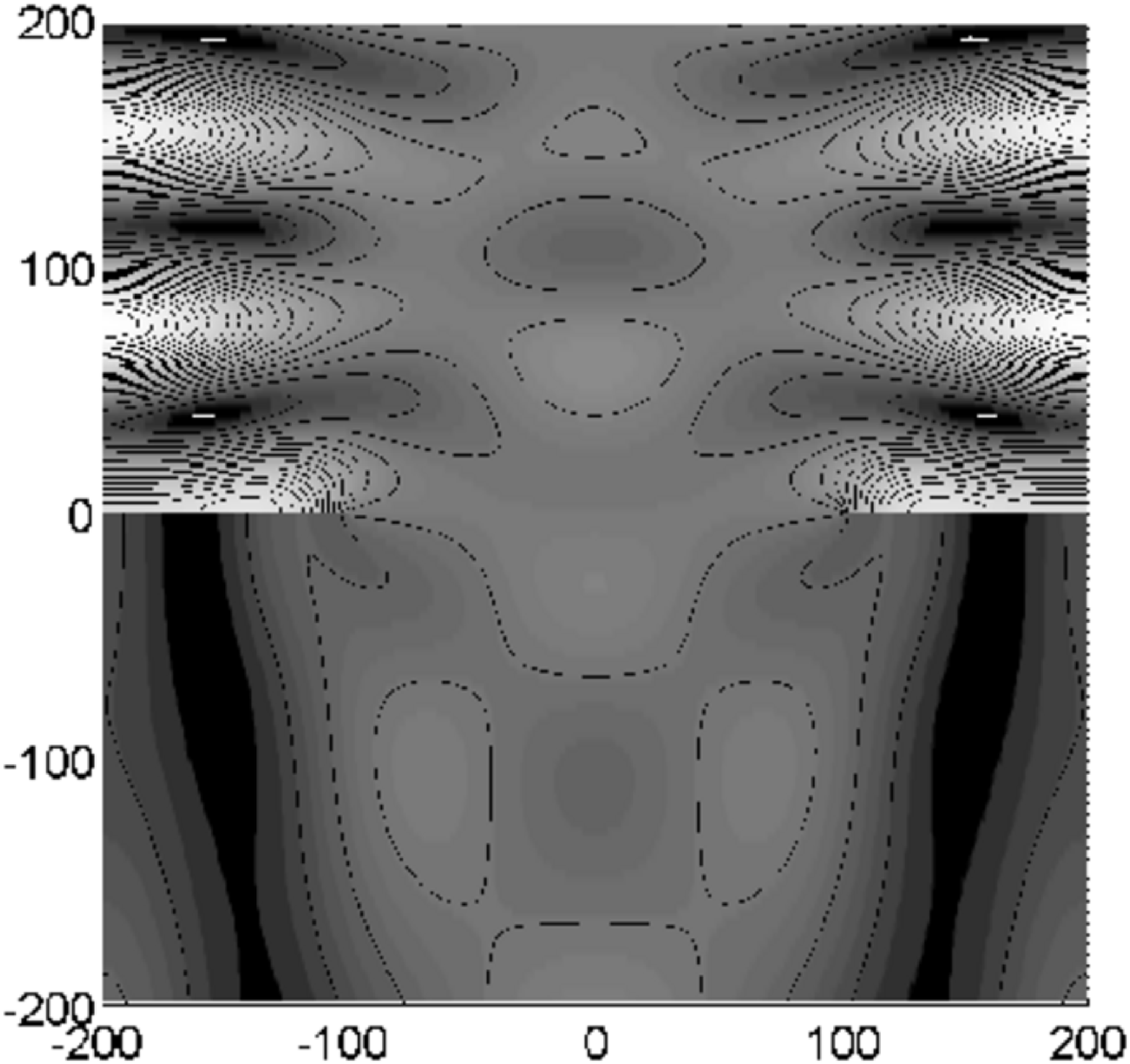}
  \caption{\label{geometry}Left: Geometry of the problem reduced
to a 2D scheme in the $x,z$ plane. Right: Intensity plot of the
superposition of incident and scattered field  (black: low intensity, white: high intensity).}
\end{figure}

The attainable absolute accuracy of the finite element solver was assessed using
an example which is close enough to lithography applications and allows 
for a fully vectorial 2D quasi-analytic solution by means of a series expansion
with proven convergence. The geometry is depicted in Fig.~\ref{geometry}. It consists of 
dense lines and spaces in an infinitely thin membrane mask of perfectly 
conducting metal embedded in free space. The metal layer covers the $xy$--plane, 
the strips of the grating are oriented in $y$-direction. 
The grating is illuminated from the top by a TM polarized plane wave, 
i.e., $H_{y,\mathrm{in}}=\exp(-ik_z z - i \omega t)$. 
 Geometry plus incident
wave imply the following properties: 
(A) $H_y$ does not depend on the
$y$-direction which allows us to write $H_y = H_y(x,z)$,  
Maxwell's equations separate for
the $H_y$-component yielding a Helmholtz equation, 
(B) the field is
periodic in $x$ with period $2L$,  
(C) the condition $H_y(x,0)=1$ holds true inside the gap, $|x| < a$, 
whereas the condition  $\partial_n H_y(x,0) = 0$ holds true
outside this gap, $a < |x| < L$. 
Altogether, the time harmonic Maxwell's equations for the
magnetic field component $H_y$ reduce to 
\begin{eqnarray*}
  \label{eq:helmholtz2}
  \partial_{xx} u + \partial_{zz} u + k^{2} H_y &=& 0 
  \quad \mbox{ for } (x,z)\in [-L,L] \times \mathbb{R}^{-}
  \label {HH}\\
  H_y(-L,z) &=& H_y(L,z) \qquad \mbox{ (periodicity in x) }
  \\
  \partial_{n} H_y(x,0) &=& 0  \quad \mbox{ for } a \le |x| \le L
  \\
  H_y(x,0) &=& 1  \quad \mbox{ for } |x| \le a
  \\
\lefteqn{+ \mbox{ radiation condition for the scattered field.}} \hspace{7em} &&
\end{eqnarray*}
\begin{figure}[thbp]
  \centering \includegraphics[width =
  0.6\textwidth]{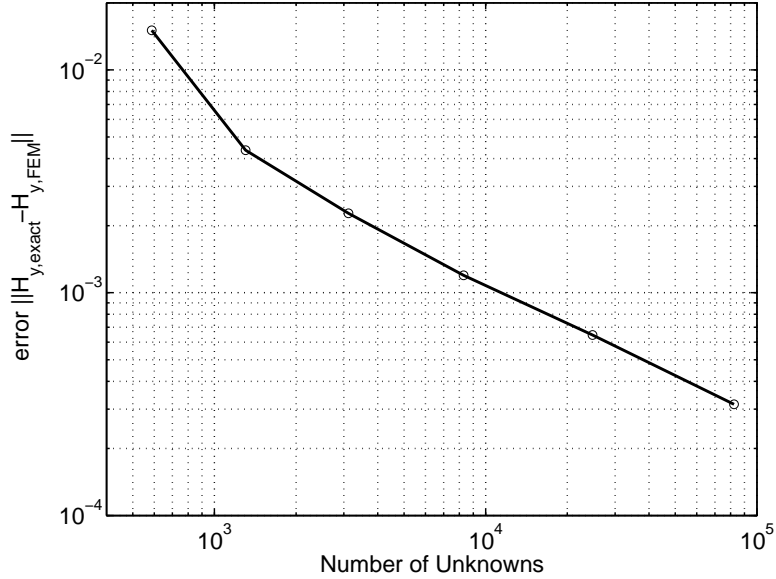}
  \caption{\label{convergence} Convergence of the FEM approach 
  towards the analytic solution.
  {\footnotesize (See original publication for images with higher resolution.)}
}
\end{figure} 
Periodicity in $x$ justifies the Fourier expansion
$
H_y(x,z) = \sum_{-N}^{N-1} c_{n}(z) \exp(in\pi x/L)
$.
Inserted into (\ref{HH}) we obtain 
 the ordinary differential equation
$c_{n}''(z) = ((n\pi/L)^{2} - k^{2}) c_{n}(z) $ for the unknown coefficients $c_{n}(z)$.
It has the general  solution  
\begin{eqnarray*}
c_{n}(z) = \left\{ 
  \begin{array}[c]{c}
    a_{n} \exp(\sqrt{(n\pi/L)^{2}-k^{2}}z) + b_{n} \exp(-\sqrt{(n\pi/L)^{2}-k^{2}}z) 
    \mbox{ for } k^{2} < (n\pi/L)^{2}
    \\
    a_{n} \exp(i \sqrt{k^{2}-(n\pi/L)^{2}}z) + b_{n} \exp(-i\sqrt{k^{2}-(n\pi/L)^{2}}z) 
    \mbox{ for } (n\pi/L)^{2} < k^{2}    
  \end{array}
\right.  \quad .
\end{eqnarray*}
The coefficients $a_n,b_n$ have to be chosen such that a proper radiation
condition holds true. This implies $b_{n}=0$ since we firstly require
decaying solutions in case of  $(n\pi/L)^{2}>k^{2}$ and $z \longrightarrow
-\infty $, and secondly, outgoing solutions in case of
 $(n\pi/L)^{2}<k^{2}$
 and $z\longrightarrow -\infty $. Based on the
ansatz for $H_y$, the derived explicit solutions for the coefficients
$c_n$ in terms of coefficients $a_n$, an uniform sampling, 
$x_n=-L+n L/N, n=-N,\ldots,N-1$, we obtain $2N$ equations for $2N$
unknowns $a_n$:
\begin{eqnarray*}
  H_y(x_n,0) = 1 &=& \sum_{n} a_n e^{in\pi x_n/L } \quad \mbox{ for } n
  \mbox{ such that } |x_n| \le a\\ 
  \partial_n H_y(x_n,0) = 0 &=& \sum_n
  a_n\sqrt{(n\pi/L)^2-k^2} e^{in\pi x_n/L} \quad \mbox{ for } n \mbox{
  such that } a \le |x_n| \le L\,.
\end{eqnarray*} 
The numerical solution of this system supplies the
quasi-analytic reference solution $H_y(x,0)$ with $|x| \le L$. The
reference solution was computed with $N=2^{12}$ ensuring an error less
than $10^{-6}$ with respect to the mean quadratic error,
$||H_{y,N}-H_{y,\mathrm{exact}}||:=1/N\sqrt{\sum_n
\left(H_{y,N}(x_n)-H_{y,\mathrm{exact}}(x_n)\right)^2}$,  on the
interval $-L \le y\le L$ at $z=0$. 
The reference solution was then compared to FEM solutions with
increasing number of unknowns, where exactly the geometry depicted in
Fig.~\ref{geometry} was used. 
The convergence
results are presented in Fig.~\ref{convergence}. The first FEM
solution was obtained for 587 unknowns yielding an error $1.506 \cdot
10^{-2}$ , the last solution was computed with $82\,303$ unknowns
yielding an error $3.158 \cdot 10^{-4}$.  
This proves that the FEM solver JCMharmony produces simulation results which 
converge to the exact solution of the scattering problem. 
The current implementations of the other solvers did not allow to 
examine this example with these.

\section{Conclusions}
\label{conclusions}
We have benchmarked a FEM solver for mask simulations against 
two other rigorous methods (FDTD and waveguide method).
The FEM solver allowed to reach high accuracies which were not 
accessible with the other methods. 
The computation time for phase mask simulations at moderate 
target accuracies was up to two orders of magnitude lower when 
using the FEM solver compared to the other methods. 
Further we have performed a benchmark of the FEM solver against 
an analytically accessible problem, and we have shown the wide range of 
applications of the solver by examining several typical simulation problems. 

\acknowledgements
We acknowledge financial support by the Ministry of 
Education and Research of the Federal Republic of Germany 
(Project No. 01M3154D), by the DFG Research Center {\sc Matheon}, and by
the priority programme SPP 1113 of the Deutsche Forschungsgemeinschaft, 
DFG.

\bibliography{/home/numerik/bzfburge/texte/biblios/phcbibli,/home/numerik/bzfburge/texte/biblios/group05,/home/numerik/bzfburge/texte/biblios/lithography}   
\bibliographystyle{spiebib}   

\end{document}